%% file: apssamp.tex
\documentclass[%
 reprint,
 amsmath,amssymb,
 aps,
]{revtex4-2}
\usepackage{comment}
\usepackage{graphicx}
\usepackage{dcolumn}
\usepackage{bm}

\usepackage{caption}
\usepackage{mathrsfs,amsmath}
\usepackage{subcaption}
\usepackage{braket}
\usepackage{physics}
\usepackage{xcolor}
\newcommand{\red}[1]{\textcolor{black}{#1}}

\begin{document}

\preprint{APS/123-QED}

\title{Topological analysis of the complex SSH model using the quantum geometric tensor}

\author{Eve Cheng\textsuperscript{1}}
\email{Wenjun.Cheng@anu.edu.au}

\author{Murray T. Batchelor\textsuperscript{1}}

\author{Danny Cocks}

\affiliation{\textsuperscript{1}Mathematical Science Institute, Australian National University, Canberra ACT 2601, Australia}

\newcommand{\Schrodinger}{Schr\"{o}dinger }
\date{\today}

\begin{abstract}
This paper presents two methods for topological analysis of the complex Hermitian Su-Schrieffer-Heeger (SSH) model using the quantum geometric tensor: Berry phase and topological data analysis. We demonstrate how both methods can effectively generate topological phase diagrams for the model, revealing two distinct regions based on the relative magnitudes of the parameters $|v|$ and $|w|$. Specifically, when $|v| > |w|$, the system is found to be topologically trivial, whereas for $|v| < |w|$, it exhibits topologically non-trivial behavior. Our results contribute to building the groundwork for topological analysis of more complicated SSH-type models.
\end{abstract}

\maketitle


\input{Intro}
\input{Method}

\input{results}

\begin{acknowledgments}
The authors thank Kate Turner for the algorithm to detect winding numbers and helpful conversations on topological data analysis in general. This work has been partially supported by Australian Research Council Grants DP210102243 {\red{and DP240100838}}.

\end{acknowledgments}

\appendix

\input{appendix}

\bibliography{apssamp}

\end{document}

%% file: Intro.tex
\section{\label{sec:level1}Introduction}
The Su-Schrieffer-Heeger (SSH) model \cite{su1979solitons,sirker2014boundary}, describing a single spinless fermion on a one-dimensional lattice with staggered hopping amplitudes, serves as the starting, yet fundamental, model for higher dimensional topological insulators. 
In particular, since the discovery of edge states in topological insulators/superconductors, the topological behaviour of the SSH model has attracted widespread attention from areas such as spintronics, magnetism and topological quantum computing \cite{li2023topological,kitaev2003fault,plugge2017majorana,li2015winding}. Subsequently the family of SSH models has grown to include many extensions of the basic SSH model, such as long range hopping \cite{li2018extended,li2014topological,rufo2019multicritical}, extended unit cells \cite{anastasiadis2022bulk,alvarez2019edge} and higher dimensional versions on the square \cite{otaki2019higher,obana2019topological} and honeycomb lattices \cite{liu2022takagi}.

The most popular and successful approach to identify the topological phases of the family of SSH models is to calculate the Berry phase \cite{berry1984quantal,bohm2003geometric}, but it has been shown that the Berry phase requires generalisation in many cases, including some very basic extensions \cite{anastasiadis2022bulk,yao2018edge,kawabata2019symmetry,tsubota2021symmetry}. 
An attempt was given very recently to explore new methods in detecting topological phases of these models using topological data analysis \cite{park2022unsupervised}.
Here our intention is to investigate this method further in terms of its performance on extended SSH models. 

Among the various extensions of the SSH model, studies involving complex hopping coefficients have gained the most interest in recent years, particularly non-Hermitian SSH models for finite systems \cite{yao2018edge,gong2018topological,tsubota2021symmetry,yu2021unsupervised,zhou2019periodic,cheng2021supervised,ryu2012analysis}. Other than the non-Hermitian models, the physical origins for the complex hopping coefficients include optomechanical systems \cite{xu2022generalized}, externally driven systems \cite{oztas2019schrieffer}, spin-orbit couplings \cite{liu2022topological}, some electrical circuits systems \cite{nunnenkamp2011synthetic,koch2010time}, photonic systems \cite{du2019phase}, and in the context of non-Abelian gauge theory\cite{nunnenkamp2011synthetic,wu2022non}. 

As the first step in experimenting with the topological data analysis approach with extended SSH models, we explore the behaviour of the SSH model with complex coupling parameters. We refer to it here as the complex SSH model. 

\subsection{The complex SSH model}

\begin{figure*}[htb] 
     \includegraphics[width=0.5\textwidth]{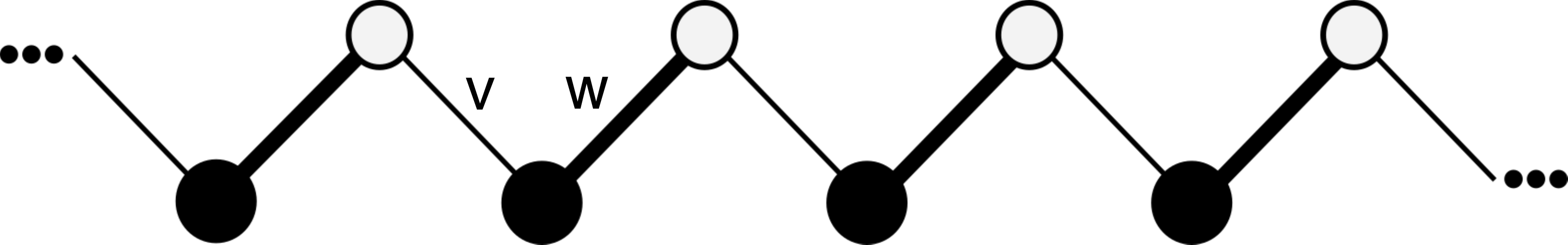}
\caption{Geometry of the one-dimensional SSH model. The labels $v$ and $w$ represent the alternating hopping coefficients.\label{ssh}}
\end{figure*}

The SSH model describes the staggered hopping of a single spinless fermion on a one-dimensional chain (see Fig.~\ref{ssh}). The Hamiltonian can be written as
\begin{eqnarray}\label{SSH1}
    H &=& \sum_{m=1}^N (v\ket{m,B}\bra{m,A} + h.c.) \nonumber \\
   &\phantom{=}& +  \sum^{N-1}_{m=1}(w\ket{m+1,A}\bra{m,B} + h.c.) \,,
\end{eqnarray}
where $A$, $B$ refer to the two sites in each unit cell, and $v$ and $w$  respectively refer to the intra- and inter-cell hopping amplitudes. \red{The hopping coefficients $v$ and $w$ are here defined as complex numbers. $\ket{m,A/B}$ refers to $\ket{m} \otimes \ket{A/B}$, which is the tensor product of the lattice and the sublattice state \cite{asboth2016short}.} When $v$ and $w$ are real, the model reduces to the original SSH model. 

\subsection{Bulk-boundary correspondence}

Bulk-boundary correspondence is a phenomenon that associates behaviour at the boundary in finite systems to those in the bulk \cite{asboth2016short,sedlmayr2018bulk,koch2020bulk,prodan2016bulk}. This correspondence is widely used because the analysis of finite systems can be drastically simplified by going to the analogous periodic system.
The advantage of periodic systems is that the translational symmetry introduced by the periodicity can decompose the Hamiltonian to much smaller invariant subspaces via the Fourier transform.  We will illustrate how this simplification can be used to study the topological behaviour of the SSH models. 

The topological behaviour of particular interest is associated with the existence of edge states, which live on the boundary of finite systems \cite{asboth2016short}. The characteristic behaviour of the edge states is a predominant peak at the boundary and
exponential decay away from the boundary. Without the establishment of the bulk-boundary correspondence, the investigation of edge states requires the system to have open boundaries.

If the bulk-boundary correspondence is established, we can safely transform the open-boundary system to a periodic-boundary system, which contains all necessary information to analyse the topological features. More specifically, with bulk-boundary correspondence, one could predict all the topological phase transition points where edge states occur in the open system from the periodic system.

In the context of the complex SSH model, which will be shown here to possess clear signs of bulk-boundary correspondence, we use the Fourier transform and Bloch's theorem to reduce the problem to a $2 \times 2$ Bloch matrix:
\begin{align}
    \ket{k} &= \frac{1}{\sqrt{N}} \sum^N_{m=1} e^{imk} \ket{m} \,,\\
    \ket{\psi_n(k)} &= \ket{k} \otimes \ket{u_n(k)} \,,
\end{align}
where $m$ is the index of the unit cell, and $k$ is the momentum vector. 
Specifically, the $2 \times 2$ matrix is of the form
\begin{align}\label{block}
\begin{bmatrix}
0 & v+w e^{-ik} \\
v^* + w^* e^{ik} & 0 
\end{bmatrix} \,.
\end{align}
With the Bloch matrices defined, the points for topological phase transitions can be identified by observing the points where the band gaps close for the Bloch matrices.

The bulk-boundary correspondence is observed in most Hermitian systems. In terms of more complicated Hermitian systems and some non-Hermitian systems, a more generalised definition of bulk-boundary correspondence is needed \cite{anastasiadis2022bulk,yokomizo2019non,xiong2018does}. 

\subsection{Topological analysis}

The goal of topological analysis is to generate a topological phase diagram. The most important and first step towards the goal is to identify the topological spaces related to the physical system such that their topological features match with the observed physical topological behaviour. Such topological spaces are not unique to any system. In fact, it is often beneficial to identify as many such spaces as possible, especially for simple systems such as the complex SSH model. Providing as many options as possible lays groundwork for future research on more complicated systems.

Here we show that there are two topological spaces that can successfully identify the topological phase transition in the physical system. The first topological space is the principal line bundles defined by the $k$-space and the $n${th} eigenvectors. The topological invariant defined on this topological space is the Berry phase. In this case, the Berry phase can also be converted into a winding number problem, which we will show in this paper. The second topological space is the eigenspace, a projective Hilbert space. The topological invariant for the topological phases is observed to be the homology class of this space. Both topological spaces and the method for computing the topology of the complex SSH model are described below.

\subsection{Outline}

In section~\ref{method}, we explain in detail how those two topological spaces can be used to generate the topological phase diagrams for the complex SSH model. We define the Berry phase and provide a detailed flowchart for an in-house Topological Data Analysis (TDA) program. In section~\ref{result}, we present the numerical results for the complex SSH model as well as the topological analysis using two methods. Finally, in section \ref{conclusion}, we summarise our findings and future directions.

%% file: Method.tex
\section{The two methods for  topological analysis \label{method}}

\subsection{The quantum geometric tensor}
The two approaches considered originate from the two aspects of the quantum geometric tensor. Here we give a brief introduction to the definition of the quantum geometric tensor and its relations to the two approaches. For detailed definitions, see Refs.~\cite{carollo2020geometry,gu2010fidelity}. 

The quantum geometric tensor, also called the Fubini-Study metric, is a gauge invariant metric defined on the projected Hilbert space.The expression for the quantum geometric tensor is 
\begin{align}
    Q_{i j \lambda} &= \bra{\partial_{i} \psi}\ket{\partial_{j}\psi} -  \bra{\partial_{i} \psi}\ket{\psi}\bra{ \psi}\ket{\partial_{j}\psi} .
\end{align}
Here $\lambda$ refers to the Hamiltonian parameters, and $i$, $j$ represent the spatial parameters for the Bloch sphere. More details on the metric and its applications are given in Ref.~\cite{gu2010fidelity}.

The two approaches here are based on the real and imaginary parts of the geometric tensor. The imaginary part of the geometric tensor can be related to the Berry curvature 2-form on the Hermitian principal line bundle (for more details see Refs.~\cite{carollo2020geometry,gu2010fidelity}). Using the Chern-Gaussian-Bonnet theorem \cite{chern1945curvatura}, the Berry curvature can be used to calculate the Berry phase, which is the topological invariant used in the first approach.  

The second approach depends on the real part of the quantum geometric tensor, the fidelity. It provides a natural metric to describe the distance between quantum states \cite{ye2023quantum} (see \ref{tdametric} for the exact expression). Recently, it was shown that fidelity can be used to directly detect topological phase transitions \cite{ye2023quantum}. Ref. \cite{park2022unsupervised} used fidelity as a metric on the projected Hilbert space and explored its geometry. Here we follow closely the approach from \cite{park2022unsupervised} to explore how the geometry behaves for a slightly different model.

\begin{figure*} 
     \includegraphics[width=0.7\textwidth]{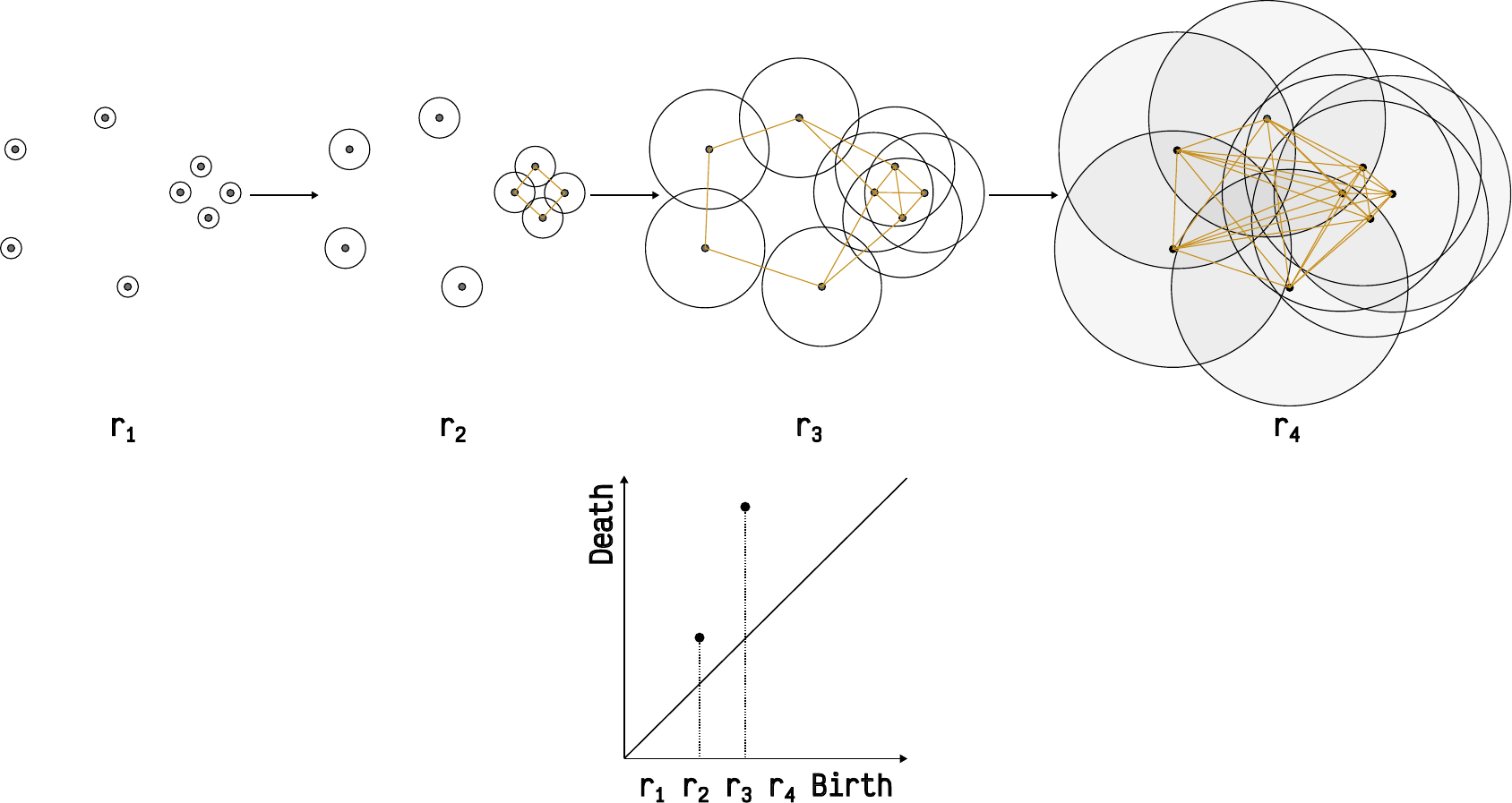}
\caption{Demonstration of the procedure of topological data analysis. The example persistence diagram here only records the 1-homology classes. Starting from discrete data points as vertices, circles of radius of $r$ are drawn. As $r$ increases and the circles intercept, edges were drawn among the vertices. A series of simplicial complexes are built during the growth of $r$ and the corresponding homology module is recorded into a persistence diagram in the form of the births and deaths of homology generators.  The two points in the persistence diagram corresponds to the two circles formed at $r_2$ and $r_3$ respectively.\label{fig:TDA}}
\end{figure*}

\subsection{Topology of the principal line bundle: Berry phase \label{berryphase}}

The topological space upon which the Berry phase was defined is the principal line bundle G: ($E, B, \pi, F$), which is associated with the Bloch matrix (\ref{block}). $B$ is the base manifold as $k \in [-\pi,\pi]$, $F$ is the fiber as the $n${th} eigenvector, $E$ is the total space and $\pi$ is the projection map. The Berry phase essentially corresponds to the holonomy class defined on the Hermitian principal line bundle \cite{simon1983holonomy,bohm2003geometric}. The expression for the Berry phase is

\begin{align}
     \gamma &= \int_c i\bra{\Psi_n(k)} \frac{\partial}{\partial k} \ket{\Psi_n(k)}dk \,,
\end{align}
where $k$ is the wavenumber, and $\Psi_n$ is the $n${th} eigenvector of the Bloch matrix, and $c$ is the curve traced by $k$ in the base manifold. In this paper, we take $n$ to be 1 to study the lowest energy band and will omit the subscript in the following. The expression inside the integral is the connection defined on the Hermitian line bundle with the assumption of adiabatic evolution of the system.

In the physical context, the Berry phase is an extra phase the state picks up while travelling in the parameter space (in this case, the $k$ space) in a closed loop, in addition to the dynamic phase factor which comes from the time-dependent \Schrodinger equation (TDSE)~\cite{bohm2003geometric}. The Berry phase cannot be gauged away, and therefore is physically significant. Below we give a brief derivation of the expression for the Berry phase from this perspective.

We start with
\begin{align}
     \ket{\Psi(t)}  \rightarrow e^{i \theta(t)} e^{i \gamma(t)} 
    & \ket{\Psi(t)}, \qquad \theta(t) = -\frac{1}{\hbar}\int_0^t E(t')dt' \,,
\end{align}
where $\ket{\Psi(t)}$ is the state at time $t$, $\theta(t)$ is the dynamic phase factor, and $\gamma(t)$ is the Berry phase. 

We then substitute the expression into the TDSE: 
\begin{align}
     i\hbar \partial t [e^{i \theta(t)} e^{i \gamma(t)} 
     \ket{\Psi(t)}] &= E  [e^{i \theta(t)} e^{i \gamma(t)} \ket{\Psi(t)}] \,, \\
     \frac{d \gamma}{dt} &= i \bra{\Psi(t)} \frac{\partial}{\partial t} \ket{\Psi(t)} \,.
\end{align}

The last step is the variable substitution to arrive at the expression for the Berry phase:
\begin{align}
     \frac{\partial}{\partial t} \ket{\Psi(t)} &= \frac{\partial \ket{\Psi(t)}}{\partial k} \frac{\partial k}{\partial t} \,, \\
     \gamma &= \int_c i\bra{\Psi(k')} \frac{\partial}{\partial k} \ket{\Psi(k')} \,.
\end{align}

The Berry phase as a topological invariant has been shown to be successful in most chiral Hermitian systems. It does not, however, work for all chiral Hermitian systems. For example, recent work has shown that a generalised Berry phase is required for the SSH$_4$ chain~\cite{anastasiadis2022bulk}. 

It is also well-known that the Berry phase calculation can be mapped to a winding number problem. Based on the form of the eigenvector stated in section \ref{berryresult}, it is intuitive that the winding number can be calculated through tracing the circle $\phi(k)$ with coordinates from equation \ref{winding_form}. The implementation of the winding number program is explained in Appendix \ref{windingprogram}.

The winding number or the Berry phase, in essence, observes the eigenvector in $k$-space using the Euclidean metric, and the topological behaviour lies in the detection if the origin is enveloped in the circle $\phi(k)$ traces. TDA, as we show below, shows an approach with potential for detecting more nuanced changes in the geometry of the traces of the eigenvectors with different metrics. Rather than calculating a specific topological invariant, the approach observes the geometry of the eigenspace based on a chosen metric, the fidelity. In the case of the complex SSH model, we observed that in the topologically trivial phase, the eigenvector travels in a path forward and back in the same way, while in the topological non-trivial phase the eigenvector forms a circle. Therefore, the topological change in the geometry of the eigenvector space can be captured by the change in the fundamental group of the space in the TDA approach.

\subsection{Topology of the eigenspace: topological data analysis approach \label{tdametric}}

The topological space topological data analysis (TDA) explores is the projective Hilbert space while the $n${th} ($n = 1$) state travels in the $k$ space. The projective Hilbert space is the space for $\ket{\psi_{k_i}}\bra{\psi_{k_i}}$. The metric is fidelity, also known as the Hilbert-Schmidt distance, which describes the geodesic of the states lying on the Bloch sphere,  defined as 
\begin{align}
    d(\psi_{k_i},\psi_{k_j}) = \sqrt{1 - |\bra{\psi_{k_i}} \ket{\psi_{k_j}}|^2} \,.
\end{align}
This approach was introduced in Ref.~\cite{park2022unsupervised}. We have developed our own in-house program to produce the results reported in this article, and plan to make the program open-sourced. 

The topological invariant coming from this projective Hilbert space is its homology class. The idea of this second approach is to use TDA to analyse the homology classes of this space numerically. The next section provides a brief introduction to TDA.

\subsubsection{Topological data analysis}

Most briefly stated, TDA is a technique that studies a data set using topology (for full details see Ref.~\cite{dey2022computational}). 
TDA takes a data set as input and calculates the persistence modules of the data set. The basic procedure for calculating $n${th} homology modules is as follows: starting from each discrete point, which is considered a vertex in a simplicial complex, an $n$-dimensional ball of radius $r$ is defined centering the vertex. As the $n$-dimensional balls start to intercept with one another, an edge is drawn between the vertices. As $r$ grows larger, a series of simplicial complexes are hence defined. We record the set of simplicial complexes at the point of change in terms of homology as $\{X_0, X_1, \ldots, X_n\}$, with $X_0$ being the trivial homology with $r$ tending to infinity. The corresponding homology module is as follows:
\begin{align}
H_p \mathscr{F}: 0 = H_p(X_0) \rightarrow H_p(X_1) \rightarrow ... \nonumber\\ \rightarrow H_p(X_n) = H_p(X) \,,
\end{align}
where $\mathscr{F}$ is the space filtration, and the right arrows represent the inclusion homomorphisms. 

The persistence modules can be recorded in persistence diagrams. The persistence diagrams record the births and deaths of $n$-homology classes as $r$ grows larger. An example of how a persistence diagram can represent the homology module is shown in Fig.~\ref{fig:TDA}.

A distance metric can be defined in the persistence diagrams space, which is helpful for identifying clusters that correspond to different topological phases. The distance metric between persistence diagrams used here is the Wassterstein distance defined by \cite{dey2022computational}
\begin{align}\label{Wass}
    d(Dgm(\mathscr{F}_f), & Dgm(\mathscr{F}_g)) = \nonumber\\ 
    & \inf \limits_{x \in \pi} \left[\sum_{x \in Dgm(\mathscr{F}_f)} (|x - \pi(x)|_q)^q \right]^{1/q} \,,
\end{align}
where $Dgm(\mathscr{F})$ refers to the persistence diagrams, and $\pi$ is the bijection between two persistence diagrams where all points in the persistence diagrams are considered to be points in $R^2$.

\subsubsection{Implementation for the TDA approach \label{imple}}

The overall implementation of the TDA approach can be summarised in a series of steps:\\

\noindent
\textit{Step 1:} define the input parameter space for the Bloch matrix.\\
\textit{Step 2}: Calculate the persistence diagram for each set of parameters.\\
\textit{Step 3:} Conduct cluster analysis on the graph of persistence diagrams.\\
\textit{Step 4:} Output the topological phase diagram.\\

A persistence diagram (step 2) is also calculated in a series of steps:\\

\noindent
\textit{Step PD1:} For each set of parameters, find the optimised list of $k$ samples from $k$ space. \\
\textit{Step PD2:} Calculate all eigenvectors corresponding to the list of $k$. \\
\textit{Step PD3:} Compute the distance matrix using the fidelity as the distance metric between eigenvectors.\\
\textit{Step PD4:} Compute the persistence diagram. 

In the above, Step 1 and PD2 are quite self-explanatory. More details associated with the other steps are given as follows.

\subsubsection{Step PD1: Optimisation of the sampling of $k$-space}

An optimisation procedure is implemented to find a list of $k$ values from $[-\pi, \pi]$ such that the eigenvectors spread in a uniform manner in the eigenspace. This step can ensure the uniform coverage and reduce the noise for the clustering analysis in Step 3.

Step PD1 optimises the sampling of $k$-space by minimising a loss function (details of the numerical implementation are given Appendix~\ref{num detail}). Here we define our loss function, which is effective and relatively simpler compared to the loss function used in Ref.~\cite{park2022unsupervised}. The optimisation is not unique and many different loss functions can be defined, as long as they work to form a representative sample of the eigenvector space. If we have the pairwise distances between eigenvectors as the set $L_d = \{d(\psi_{k_0},\psi_{k_1}),d(\psi_{k_1},\psi_{k_2}),\ldots,d(\psi_{k_{n-1}},\psi_{k_n})\}$, where $n$ is the number of samples taken from $k$-space, the loss function is defined as
\begin{align}
    d_{\text{loss}} = \text{std}(L_d)/\text{mean}(L_d) \,,
\end{align}
where std and mean refer to the standard deviation and the mean of $L_d$. This choice of loss function serves to avoid ``clumping" of the eigenvectors together to spread them across the available space.

An example of the effect of the optimisation is shown later in Fig.~\ref{fig:bigresult}(a)-(d). The implications of the result will be discussed in the results section.

\subsubsection{Steps PD3-PD4: generating all the persistence diagrams for all sets of parameters in the parameter space}

 The inputs required to generate the persistence diagrams are pair-wise distances between the eigenvectors stored in distance matrices (where $D_{ij} = d(\psi_{k_i},\psi_{k_j})$)  (for details of the numerical implementation see Appendix~\ref{num detail}). After repeating for all sets of parameters for steps PD3 and PD4, all distance matrices for the eigenspaces and the corresponding persistence diagrams are calculated.

\subsubsection{Step 3: clustering analysis of persistence diagrams}

Once all the persistence diagrams for all sets of parameters are collected, we define a weighted graph G = ($V, E, w$) for the persistence diagrams, where $V$ and $E$ are the sets of vertices and edges respectively, and $w$ is a mapping from the edges to their corresponding weights: $w: E \rightarrow R$. The vertices $V$ are the persistence diagrams corresponding to different parameter sets, and the edge weights are defined with the distance between the persistence diagrams:
\begin{align}\label{edgeweight}
    w_{ij} = \exp{-W_2(PD_i,PD_j)/\overline{W_2}} \,,
\end{align}
where $W_2$ refers to the Wasserstein distance (\ref{Wass}) between persistence diagrams with $q$ set to 2. Only the 1-homology was included in the persistence diagrams. $\overline{W_2}$ refers to the average Wassterstein distance among all persistence diagrams.

Since the different topology within each region in the parameter space will be captured by the persistence diagrams, clustering analysis of the graph of persistence diagrams will give us the topological phase diagram directly. For the clustering analysis, an unsupervised spectral clustering method is used~\cite{von2007tutorial}. In the same way as Ref.~\cite{park2022unsupervised}, we define a random walk Laplacian as 
\begin{align}
    L = 1 - D^{-1}W \,,
\end{align}
where $W$ is the adjacency matrix where $W_{i,j} = w_{ij}$ and $D$ is a diagonal matrix, where $D_{ij} = \sum_j w_{ij}$. 

The clustering analysis starts by taking lowest eigenvalues and plotting the corresponding eigenvectors. The lowest eigenvalues of the Laplacian signal the existence of clusters in the graph, and the space where clustering analysis takes place involves the eigenvectors corresponding to those eigenvalues. Following Ref.~\cite{park2022unsupervised}, $k$-means clustering was used to detect the clusters on the embedding of the eigenvectors to $R^n$.

%% file: results.tex
\begin{figure*}
\centering
\captionsetup[subfigure]{justification=centerlast}
    \begin{subfigure}{0.45\linewidth}
        \includegraphics[width=\textwidth]{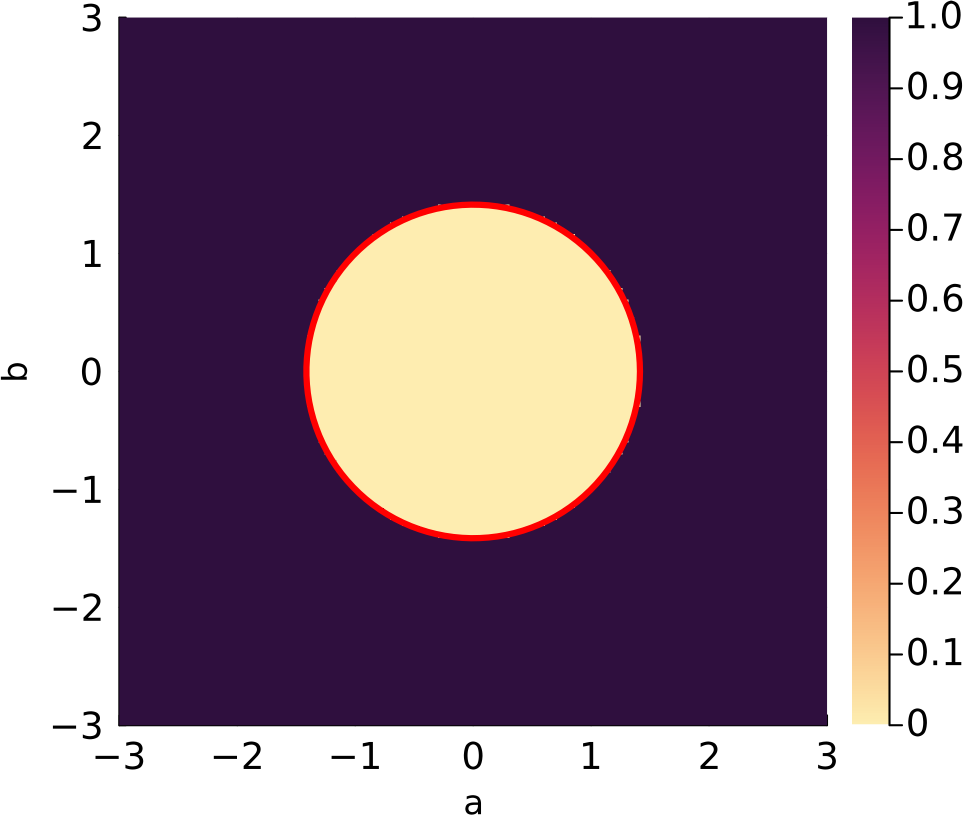}
        \subcaption{Berry phase}
    \end{subfigure}
    \hspace{0.08\linewidth}
    \begin{subfigure}{0.45\linewidth}
        \includegraphics[width=\textwidth]{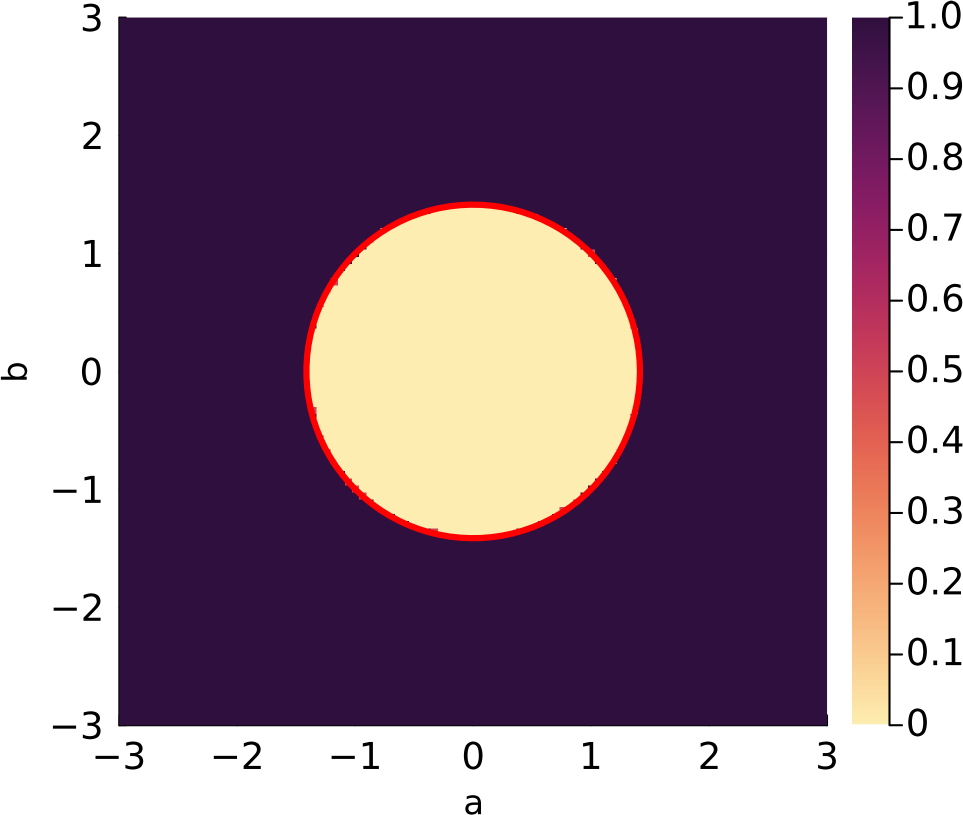}
        \subcaption{Winding number}
    \end{subfigure}
\caption{(a) Numerical result for the Berry phase with $v = 1 + i$ and $w = a + b \, i$. The yellow and purple regions correspond to topological and trivial respectively. The red line is a line plot for a perfect circle with radius $\sqrt{2}$ as a guide for the eye. (b) Numerical result for the winding number with the same set up as Berry phase\label{Berry}}
\end{figure*}

\begin{figure*}
\captionsetup[subfigure]{justification=centerlast}
  \begin{subfigure}[h]{0.31\textwidth}
     \includegraphics[width=\textwidth]{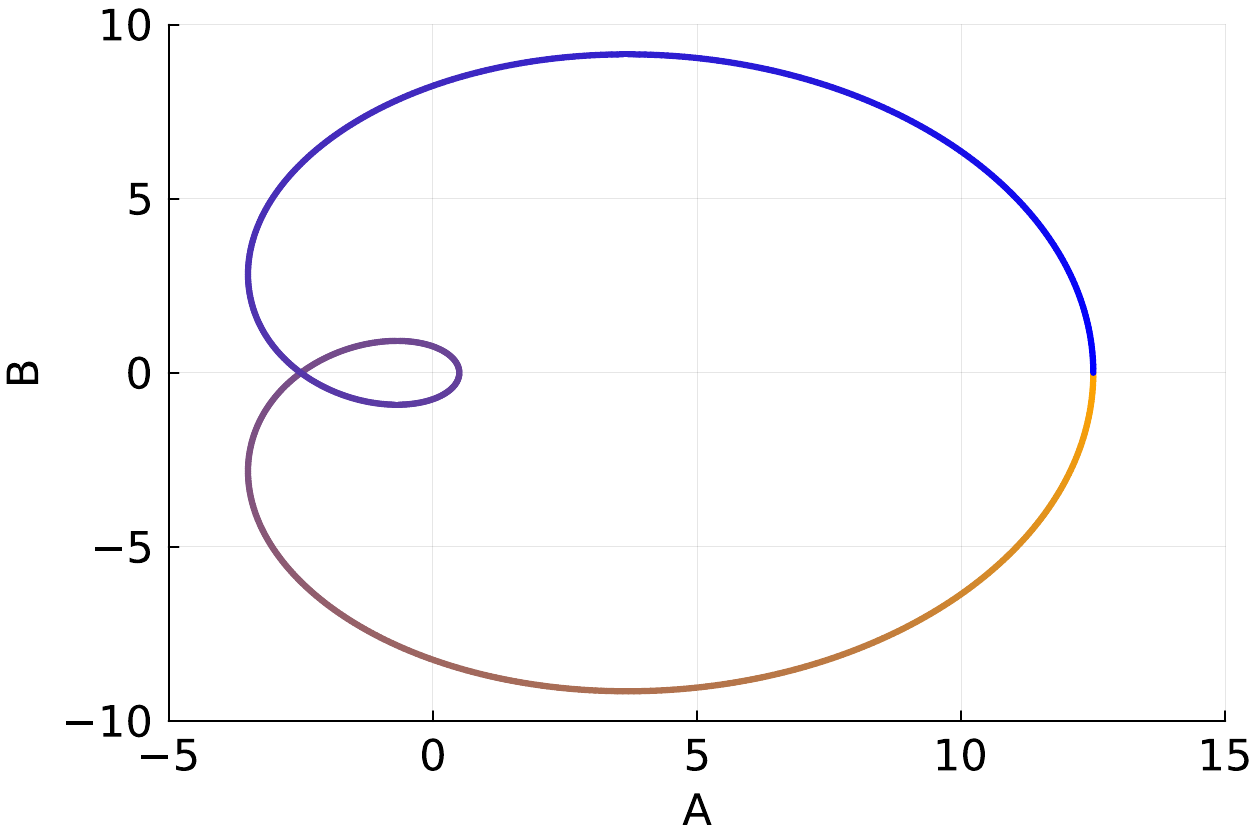}
     \caption{}
  \end{subfigure}
    \hspace{10pt}
  \begin{subfigure}[h]{0.31\textwidth}
     \includegraphics[width=\textwidth]{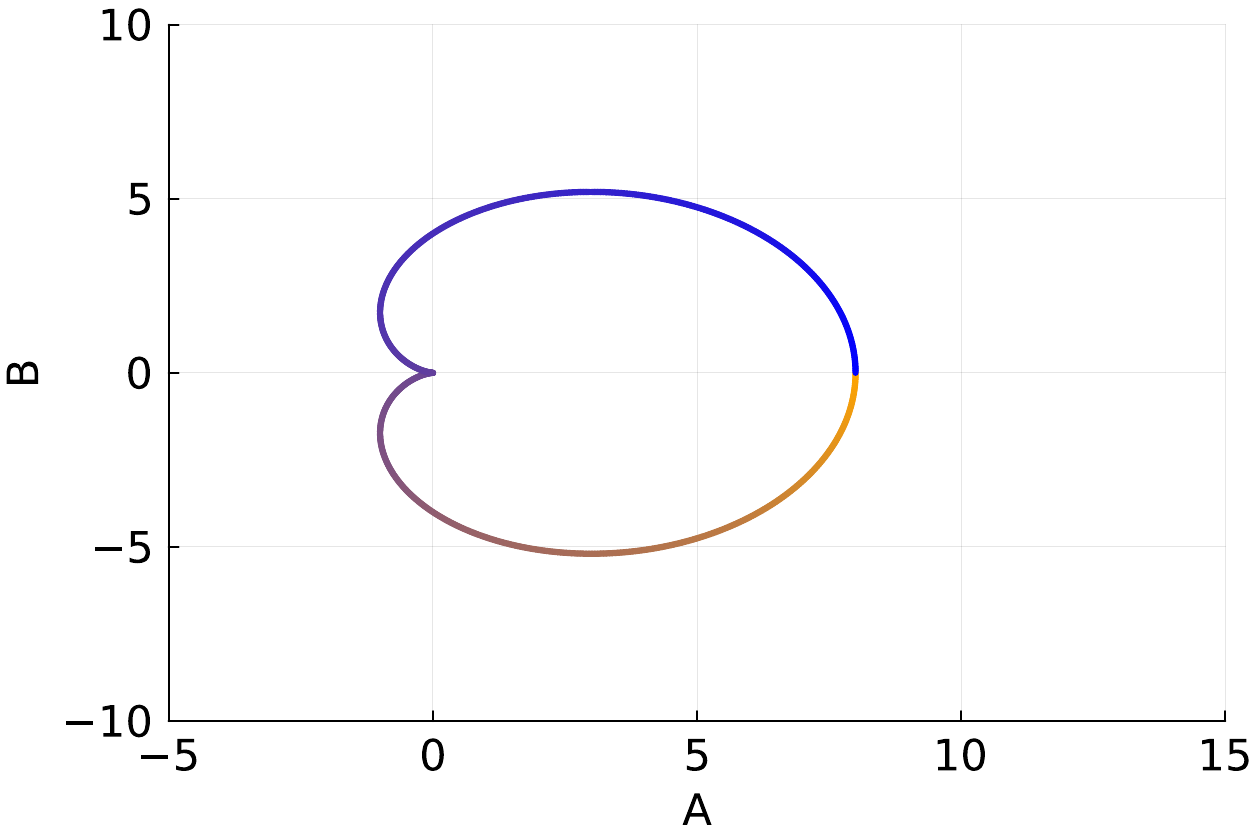}
     \caption{}
  \end{subfigure}
    \hspace{10pt}
 \begin{subfigure}[h]{0.31\textwidth}
     \includegraphics[width=\textwidth]{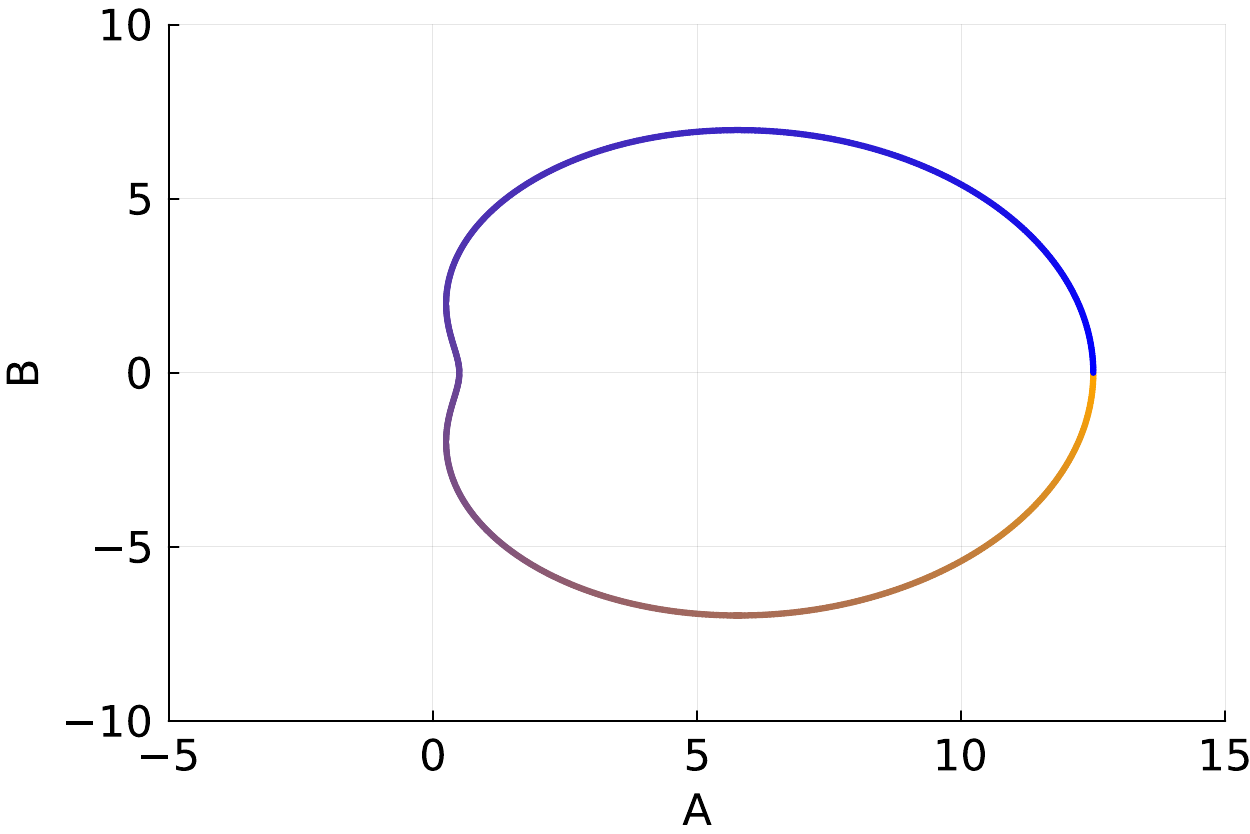}
     \caption{}
  \end{subfigure}
\caption{Behaviour of the trace $\phi(k)$, where we define 
$A=|v|^2 + 2(v_Rw_R + v_I w_I)\cos k + |w|^2 \cos 2k$ and 
$B=-(v_Rw_R + v_I w_I + |w|^2\cos k) \sin k$ from Eqn.~(\ref{winding_form}), where $\tan(2\phi(k)) = B/A$. (a) $v = 1 + i$, $w = 1.5 + 1.5i$ (b) $v = 1 + i$, $w = 1 + i$ (c) $v = 1.5 + 1.5i$, $w = 1 + i$.}
\label{winding_fig}
\end{figure*}

\section{Results \label{result}}
In this section we explore the topological features of the complex SSH model. The complex extension of this model, unsurprisingly, possesses all the topological behaviour observed in the original SSH model as well as the bulk-boundary correspondence. We have found that there are two topological phases for the complex SSH model, being $|v| < |w|$ and $|v| > |w|$. If $v$ and $w$ are real, as in the case of the original SSH model, the two topological phases reduce to $v < w$ and $v > w$ \cite{asboth2016short}. 

Analysis of both the topology of the principal line bundle and the eigenspace are shown here. 
This includes numerical results for the Berry phase and the results of the TDA approach.
\begin{figure*}[t]
\captionsetup[subfigure]{justification=centerlast}
  \begin{subfigure}[t]{0.23\textwidth}
     \includegraphics[width=\textwidth]{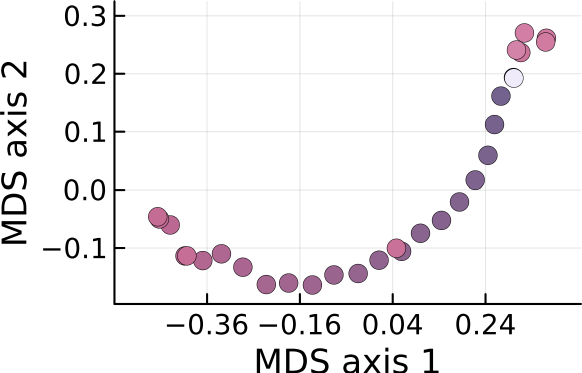}
     \caption{}
  \end{subfigure}
    \hspace{1pt}
  \begin{subfigure}[t]{0.23\textwidth}
     \includegraphics[width=\textwidth]{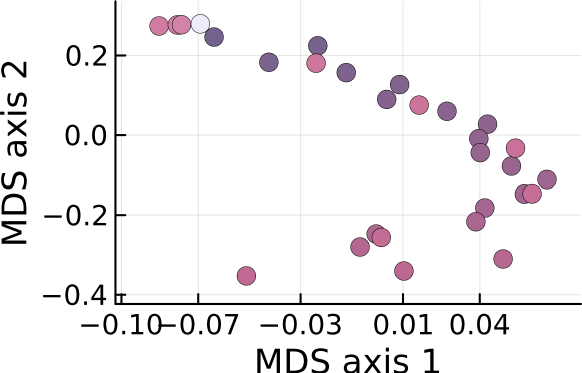}
     \caption{}
  \end{subfigure}
    \hspace{1pt}
   \begin{subfigure}[t]{0.23\textwidth}
     \includegraphics[width=\textwidth]{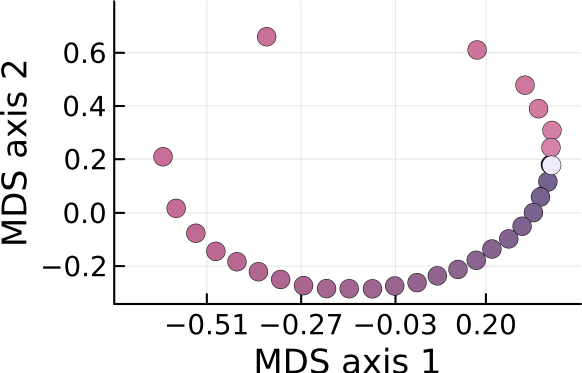}
     \caption{}
  \end{subfigure}
    \hspace{1pt}
  \begin{subfigure}[t]{0.23\textwidth}
     \includegraphics[width=\textwidth]{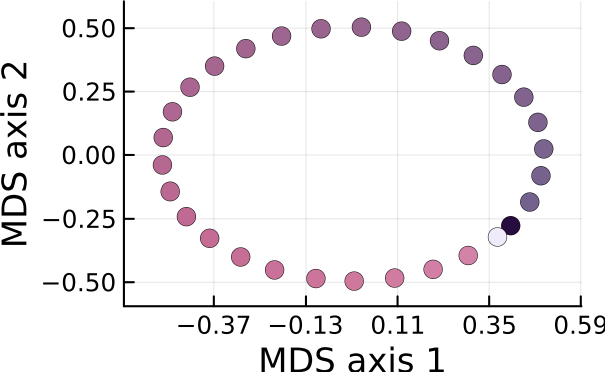}
     \caption{}
  \end{subfigure}
    \hspace{1pt}   

\begin{subfigure}[t]{0.23\textwidth}
     \includegraphics[width=\textwidth]{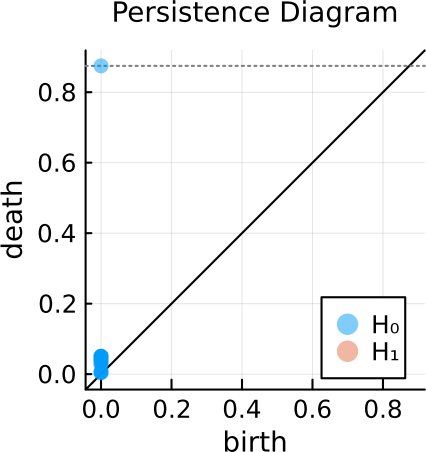}
     \caption{}
  \end{subfigure}
    \hspace{1pt}
  \begin{subfigure}[t]{0.23\textwidth}
     \includegraphics[width=\textwidth]{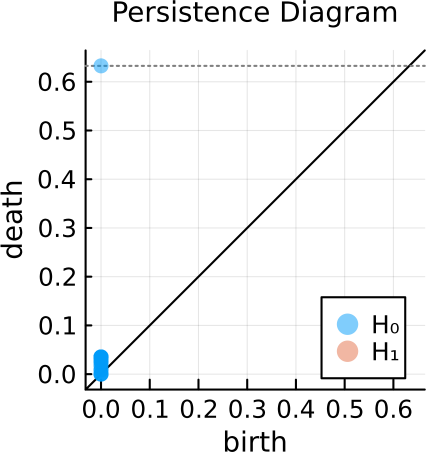}
     \caption{}
  \end{subfigure}
    \hspace{1pt}
   \begin{subfigure}[t]{0.239\textwidth}
     \includegraphics[width=\textwidth]{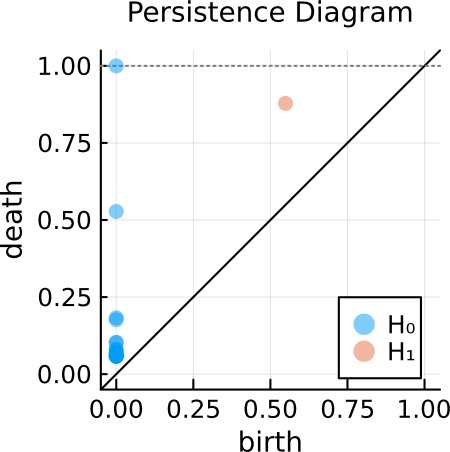}
     \caption{}
  \end{subfigure}
    \hspace{1pt}
  \begin{subfigure}[t]{0.239\textwidth}
     \includegraphics[width=\textwidth]{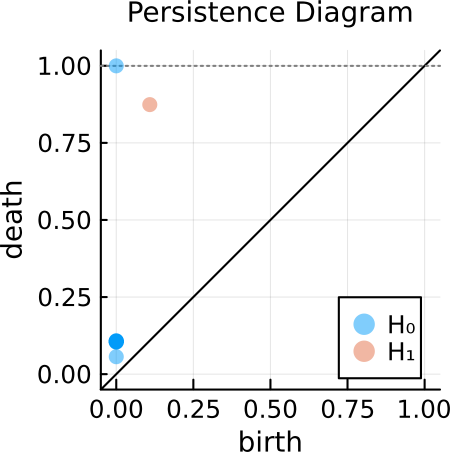}
     \caption{}
  \end{subfigure}
    \hspace{1pt}  

\caption{(a) \& (c) The multi-dimensional scaling (MDS) projection of the eigenspaces into the Euclidean space without optimisation (Step PD1 in Section~\ref{imple}) for $v = 0.8 + 1.4i$, $w = 1 + i$, and $v = 1 + i$, $w = 0.8 + 1.4 i$, respectively. (b) \& (d) The corresponding MDS projection of the eigenspace into the Euclidean space with optimisation for $v = 0.8 + 1.4i$, $w = 1 + i$, and $v = 1 + i$, $w = 0.8 + 1.4 i$, respectively. (e)-(h) The corresponding persistence diagrams of the eigenspaces above. $H_0$ and $H_1$ corresponds to the 0-homology and 1-homology group respectively \label{fig:bigresult}}

\end{figure*}
\begin{figure*}[t]
\captionsetup[subfigure]{justification=centerlast}
    \begin{subfigure}[t]{0.42\textwidth}
     \includegraphics[width=\textwidth]{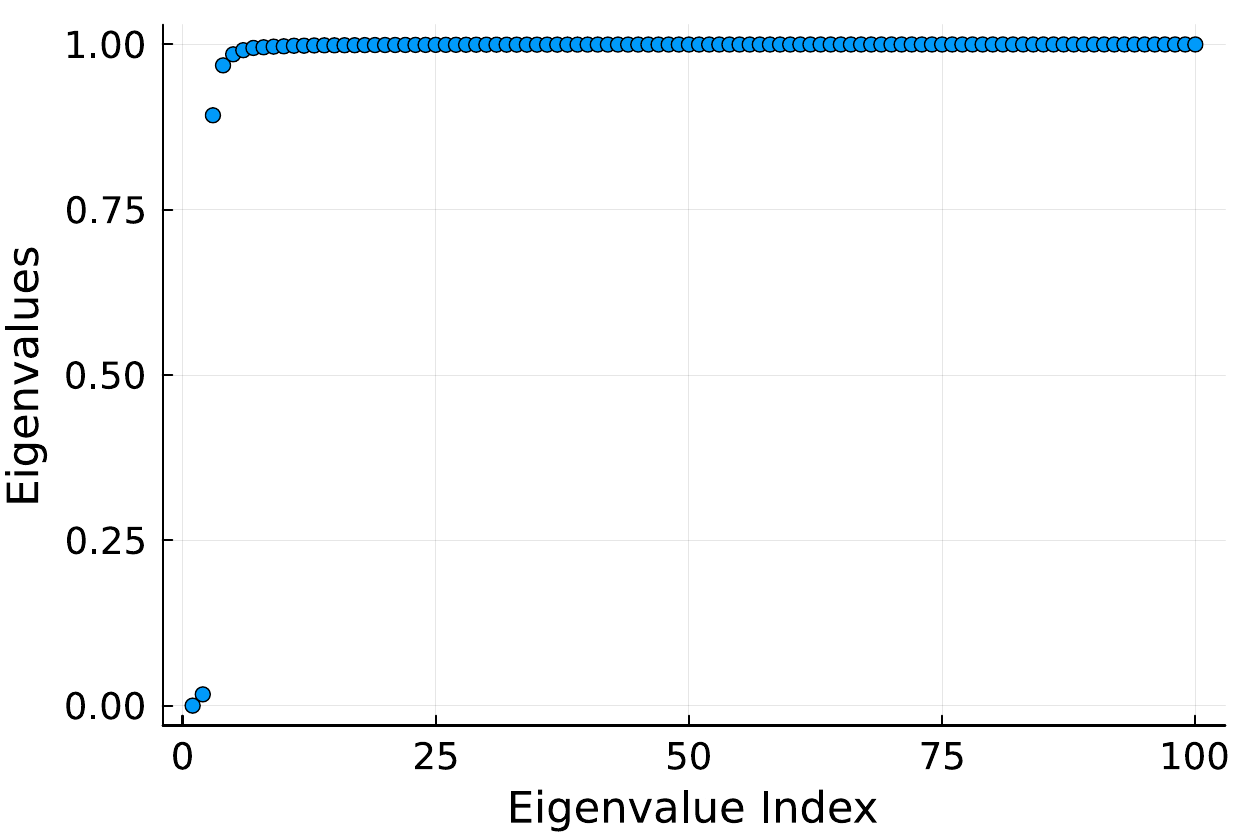}
     \caption{}
  \end{subfigure}
    \hspace{30pt}
  \begin{subfigure}[t]{0.42\textwidth}
     \includegraphics[width=\textwidth]{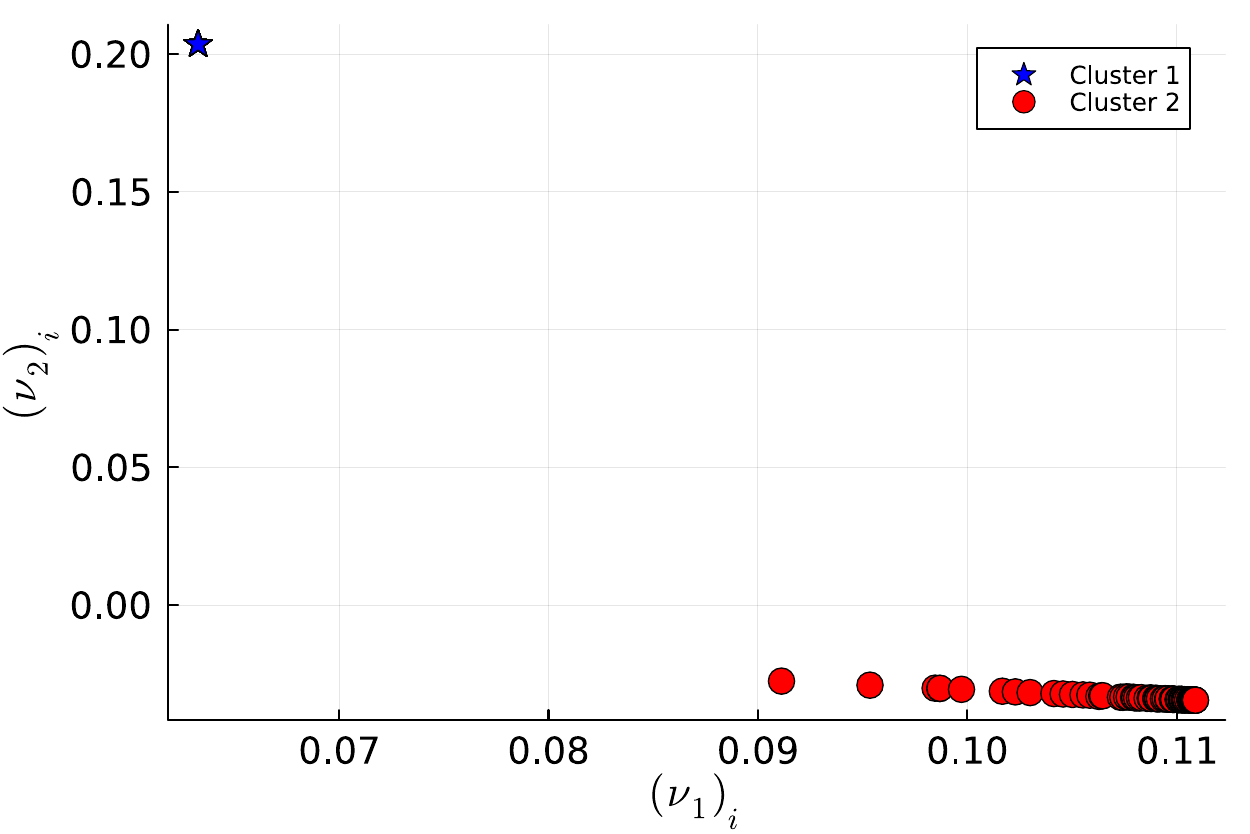}
     \caption{}
  \end{subfigure}
\caption{(a) The eigenvalues for the random walk Laplacian for the graph of persistence diagrams, with edge weights defined by  equation~(\ref{edgeweight}). Two values lower than 0.2 were considered to represent two clusters (b) The first two smallest eigenvectors plotted against each other for $k$-means cluster analysis. The blue star cluster at the top left correspond to the persistence diagrams where $|v| > |w|$ , and the red cluster correspond to the persistence diagrams where $|v| < |w|$. \label{spectral} }  
\end{figure*}
\subsection{Numerical results for the Berry phase and the winding number\label{berryresult}}

The analytical solution for Eqn.~(\ref{block}) is simply 
\begin{align}
    \ket{\pm k} = \begin{bmatrix}
\pm e^{-i\phi(k)}\\
1
\end{bmatrix} \,,
\end{align}
where 

\begin{align}
    \tan(2\phi(k)) = \frac{-(v_Rw_R + v_iw_i + |w|^2\cos k) \sin k}{|v|^2 + 2(v_Rw_R + v_iw_i)\cos k + |w|^2 \cos 2k} \,.
    \label{winding_form}
\end{align}
Here $v_R, w_R$ represent the real parts of $v$ and $w$, while $v_I, w_I$ represent the imaginary parts of $v$ and $w$. Thus the Berry phase can be calculated as

\begin{align}
     \gamma &=\frac{1}{2\pi} \int_{-\pi} ^{\pi}\frac{d\phi}{dk}dk \,.
\end{align}

The numerical calculation of the Berry phase is shown in Fig.~\ref{Berry}(a). It can be observed clearly that there are two topological phases, being inside and outside the circle. It supports our conclusion that $|v| = |w|$ is the phase transition point. When $|v| > |w|$, the complex SSH model is topological trivial with Berry phase value 0. On the other hand, when $|v| < |w|$, the complex SSH model possesses an edge state and is therefore topological non-trivial, with a non-zero Berry phase, 1.\\

The numerical results for the calculation of the winding number is shown in Fig.~\ref{Berry}(b). The detected winding numbers are 0 and 2 (because the circle winds around twice), and they need to be divided by 2 according to Eqn.~(\ref{winding_form}). The behaviour of the trace of $\phi(k)$ is shown in Fig.~\ref{winding_fig}, where \ref{winding_fig}(a), \ref{winding_fig}(b), and \ref{winding_fig}(c) shows the trace in the topologically non-trivial region, the topological phase transition point and the topologically trivial region, respectively.

\subsection{Topological data analysis of the projected Hilbert space}

Similar to the analysis given in Ref.~\cite{park2022unsupervised}, we observed that the topology of the eigenspace using fidelity as the metric changes drastically from one topological phase to another and the changes agree with the Berry phase approach. We show this change in Fig.~\ref{fig:bigresult}, and how optimisation can make that difference clearer (see Fig.~\ref{fig:bigresult}(a)-(h)). Once we are confident about the changes in the geometry of the eigenspace, we can use TDA to divide the different geometries corresponding to different topological phases into different clusters and produce the final topological phase diagram. We observed that the final phase diagram agrees with the results of Section~\ref{berryresult} (see Fig.~\ref{fig:bigresult}(i)-(j)).

We start by showing the difference in topology going from one phase to another. Two sets of parameters were chosen from regions $|v| > |w|$ and $|v| < |w|$. As shown in Fig.~\ref{fig:bigresult}(a) and (c), we see that if $|v| > |w|$, no circles can be observed from the eigenspace, while a circle is shown when $|v| < |w|$. 

It is important to note here that both Fig.~\ref{fig:bigresult}(a) and (c) are projections from the projective Hilbert space to Euclidean space through Multidimensional scaling (MDS) \cite{scikit-learn}. They are faithful representations in terms of the pair-wise distances between the data points. However, they are for visualization purposes only and the persistence diagram calculations rely on the distance matrices defined in Section~\ref{imple}. 

The change in geometry from one phase to another is shown in the corresponding persistence diagrams in Fig.~\ref{fig:bigresult}(e) - (h). In the cases where a circle is observed (see Fig.~\ref{fig:bigresult}(c) and (d)), the corresponding persistence diagrams (see Fig.~\ref{fig:bigresult}(g) and (h)) possess one 1-homology generator. In the cases where no circle is observed (see Fig.~\ref{fig:bigresult}(a) and (b)), no 1-homology generators were observed in the corresponding persistence diagrams (see Fig.~\ref{fig:bigresult}(e) and (f)).

Fig.~\ref{fig:bigresult} also shows that the optimisation step (from Section.~\ref{imple}) reduces the potential noise present in the system. From Fig.~\ref{fig:bigresult}(a) to  \ref{fig:bigresult}(b), and  \ref{fig:bigresult}(c) to  \ref{fig:bigresult}(d), we can see that the spread of the samples from the eigenspace is more uniform and the existence or non-existence of a circle is more obvious after optimisation.

Such improvement from optimisation is reflected clearly in the corresponding persistence diagrams. Overall, the persistence diagrams in Fig.~\ref{fig:bigresult}(e)
and (f) show no 1-homology generators, while  Fig.~\ref{fig:bigresult}(g) and (h) show the existence of one homology generator. The effect of the optimisation step is made clear from the comparison between Fig.~\ref{fig:bigresult}(g) and (h). While both persistence diagrams show the existence of one 1-homology generator, the 1-homology generator shown after optimisation has a longer life expectancy. This extended life of the 1-homology generator after optimisation is reflected by the relative position of the corresponding data point to the diagonal line from Fig.~\ref{fig:bigresult}(g) to (h). The data point that corresponds to the 1-homology generator was born earlier and died later. This improvement means that when we calculate the distance among persistence diagrams using Eqn.~(\ref{Wass}), the persistence diagrams that possess one 1-homology generator would have less distance among them, and the difference between the ones with and without 1-homology generators would be large. Because of this reason, the optimisation step makes the clustering process later easier and more obvious.

Fig.~\ref{spectral}(a) and (b) shows the result of the $k$-means clustering analysis of the network of the persistence diagrams (Step 3 in Section~\ref{imple}). The network of persistence diagrams was built using persistence diagrams as nodes and the distance (calculated by Eqn.~(\ref{edgeweight})) among them as distance. Since the persistence diagram gives information on the geometry of the eigenspace, the similarity between the persistence diagrams reflect the similarity between the systems correspondingly. The  clustering algorithm uses spectral graph theory to identify the existence of clusters by calculating the eigenvalues of the random walk Laplacian (see Fig.~\ref{spectral}(a)). The small eigenvalues indicate high level of clustering. We observed that there are two eigenvalues that are very close to zero, which signal two clusters in the graph. To identify which persistence diagrams live in which cluster, we plot the first two eigenvectors against each other (shown in Fig.~\ref{spectral}(b)). We chose the first two eigenvectors because there are two clusters. The $k$-means algorithm can then identify those two clusters, which we found to be the two topological phases identified by Berry phase. The blue star cluster at the top left corresponds to the region where $|v| > |w|$ and the Berry phase is zero; the red cluster at the bottom right corresponds to the region where $|v| < |w|$ and the Berry phase is 1. Because the persistence diagrams with no 1-homology have no distance among them, all points are completely degenerate. The persistence diagrams with one 1-homology generator vary in terms of the lifespan of the 
circle and therefore result in a cluster with slight variation from each other.

Therefore, it is clear that the two clusters shown in Fig.~\ref{fig:bigresult}(j) correspond to the two regions in the topological phase diagram generated using the Berry phase (see Fig.~\ref{Berry}).
\begin{figure*}
\captionsetup[subfigure]{justification=centerlast}
  \begin{subfigure}[h]{0.46\textwidth}
     \includegraphics[width=\textwidth]{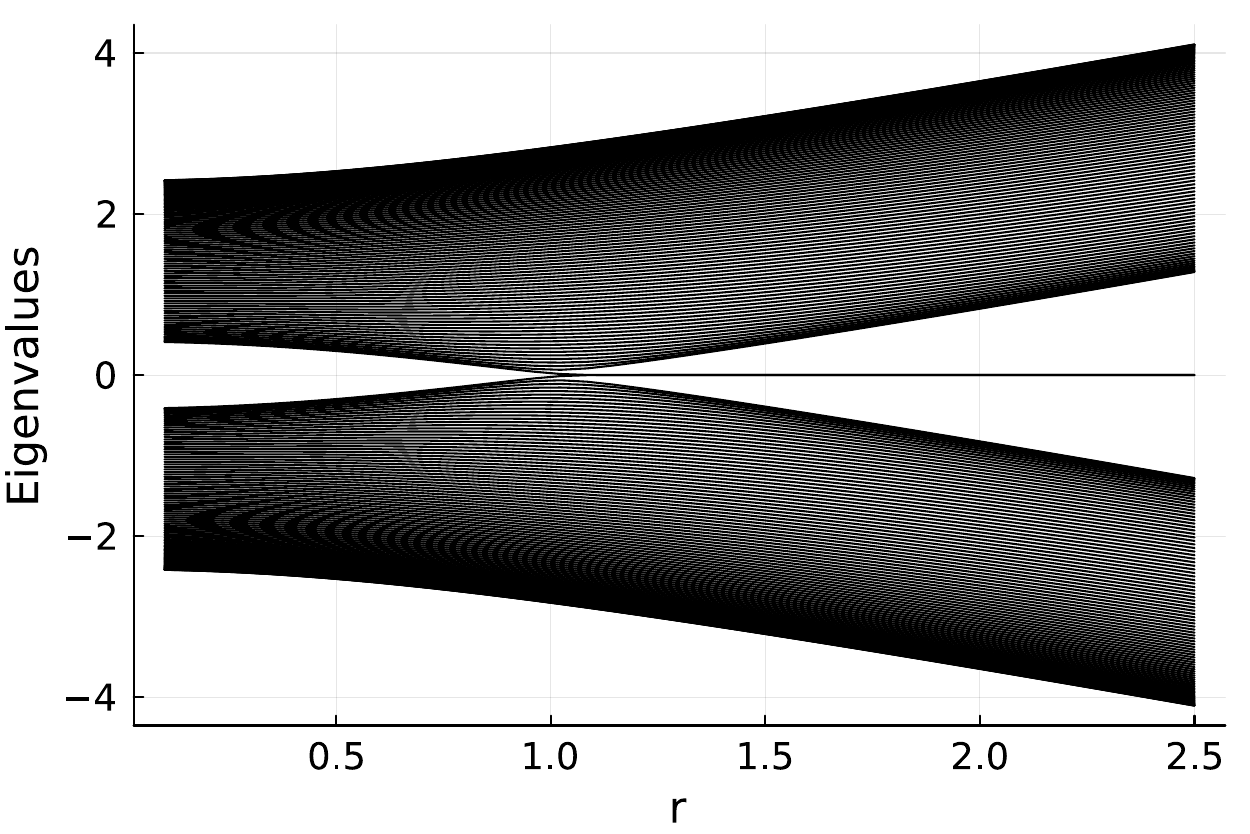}
     \caption{}
  \end{subfigure}
    \hspace{25pt}
  \begin{subfigure}[h]{0.46\textwidth}
     \includegraphics[width=\textwidth]{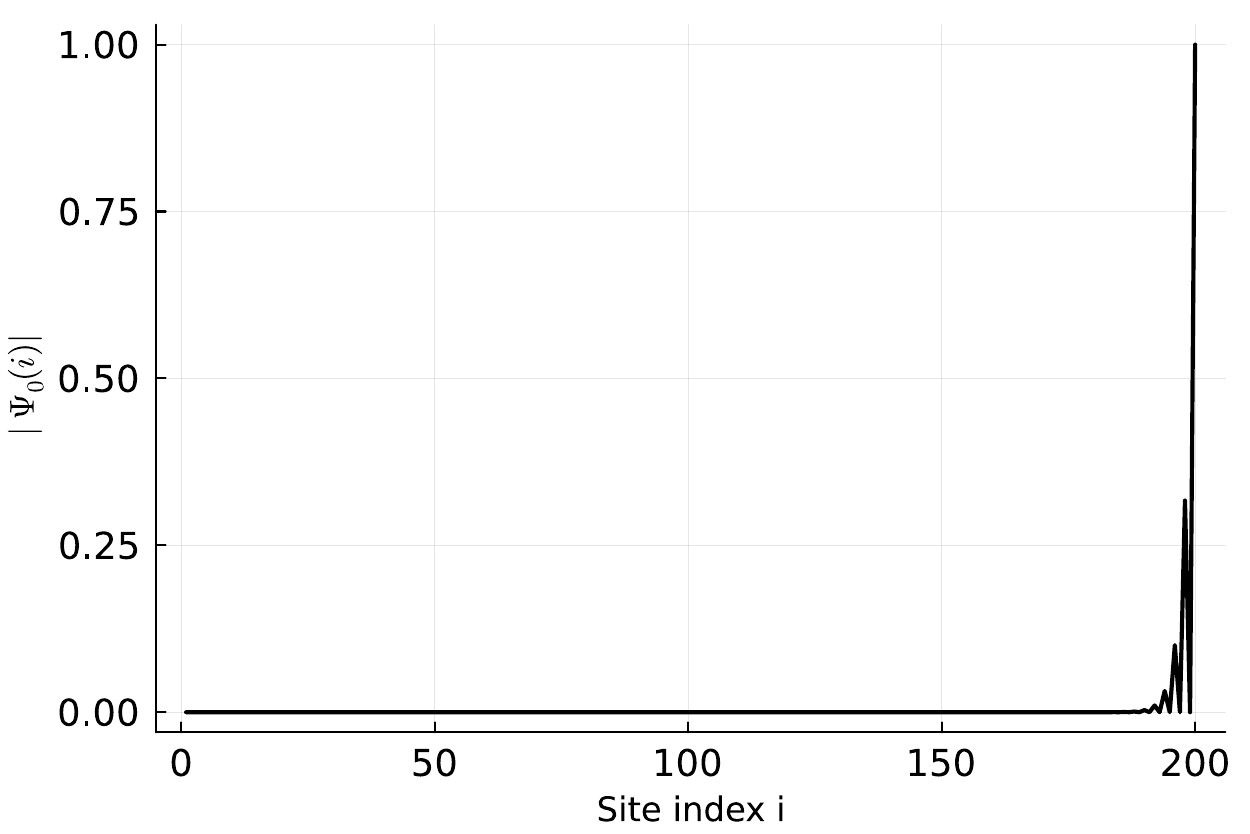}
     \caption{}
  \end{subfigure}
    \hspace{30pt}
 \begin{subfigure}[h]{0.32\textwidth}
     \includegraphics[width=\textwidth]{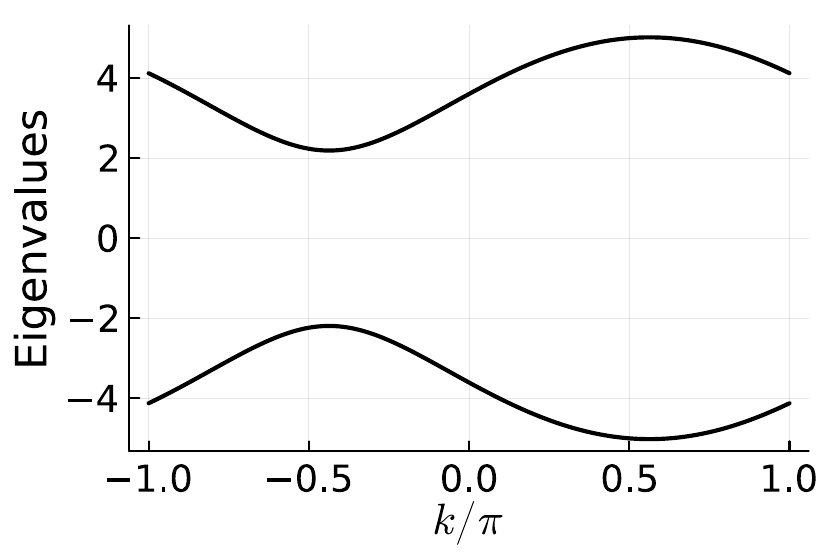}
     \caption{}
  \end{subfigure}
    \hspace{1pt}
  \begin{subfigure}[h]{0.32\textwidth}
     \includegraphics[width=\textwidth]{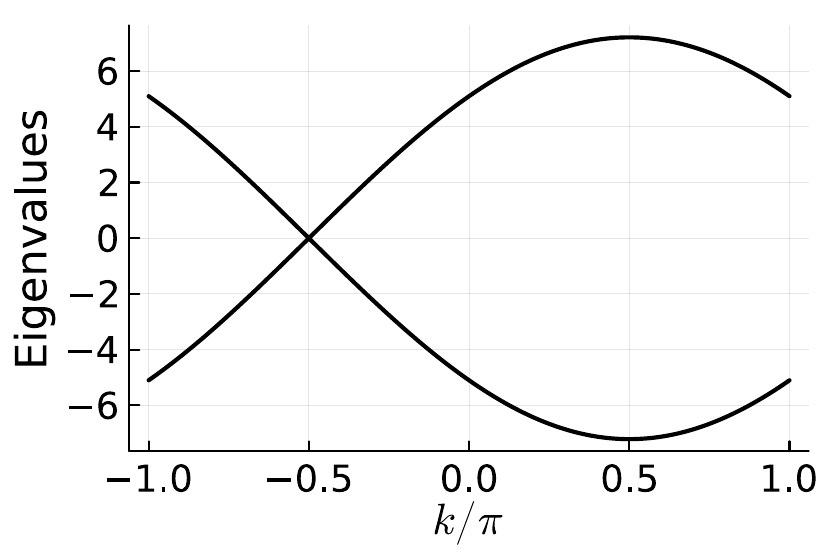}
     \caption{}
  \end{subfigure}
    \hspace{1pt}
   \begin{subfigure}[h]{0.32\textwidth}
     \includegraphics[width=\textwidth]{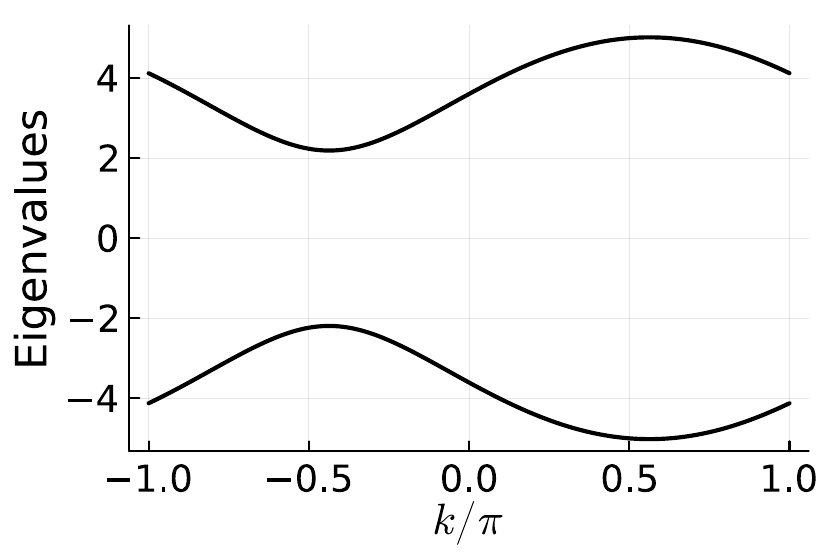}
     \caption{}
  \end{subfigure}
\caption{(a) The eigenspectrum of the complex SSH model (even number of sites) with $v = 1 + i$ and $w = 1 + r \, i$, for $N = 200$. The phase transition occurs at $r = 1$. When $r > 1$, $|v| < |w|$, and a zero energy state can be observed. The edge states can be observed in zero energy eigenvectors $\psi_0$ when $|v| < |w|$. For example, (b) shows the ground state eigenvector where $v = 1 + 1.5 i$ and $w = 2 + 2.5 i$ (c)-(e) The eigenspectrum for the Bloch Hamiltonian for ($v$,$w$) values (2 + 3 i, 1 + i), (1 + i, 2 + 3 i), (2 + 3i, 3 + 2i),  respectively.\label{fig:numresult}}
\end{figure*}
The following section shows the physical significance of the phase diagram produced by the two topological analysis approaches.
\begin{figure*}
\captionsetup[subfigure]{justification=centerlast}
  \begin{subfigure}[h]{0.32\textwidth}
     \includegraphics[width=\textwidth]{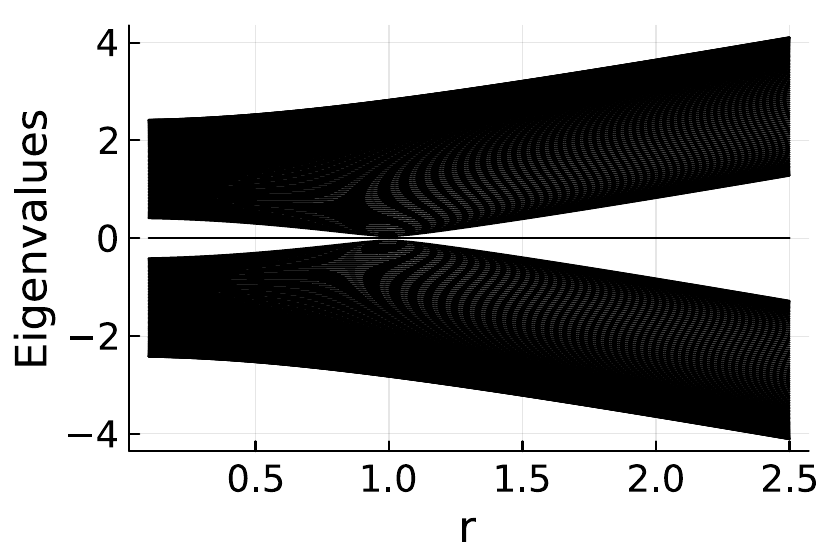}
     \caption{}
  \end{subfigure}
    \hspace{2pt}
  \begin{subfigure}[h]{0.32\textwidth}
     \includegraphics[width=\textwidth]{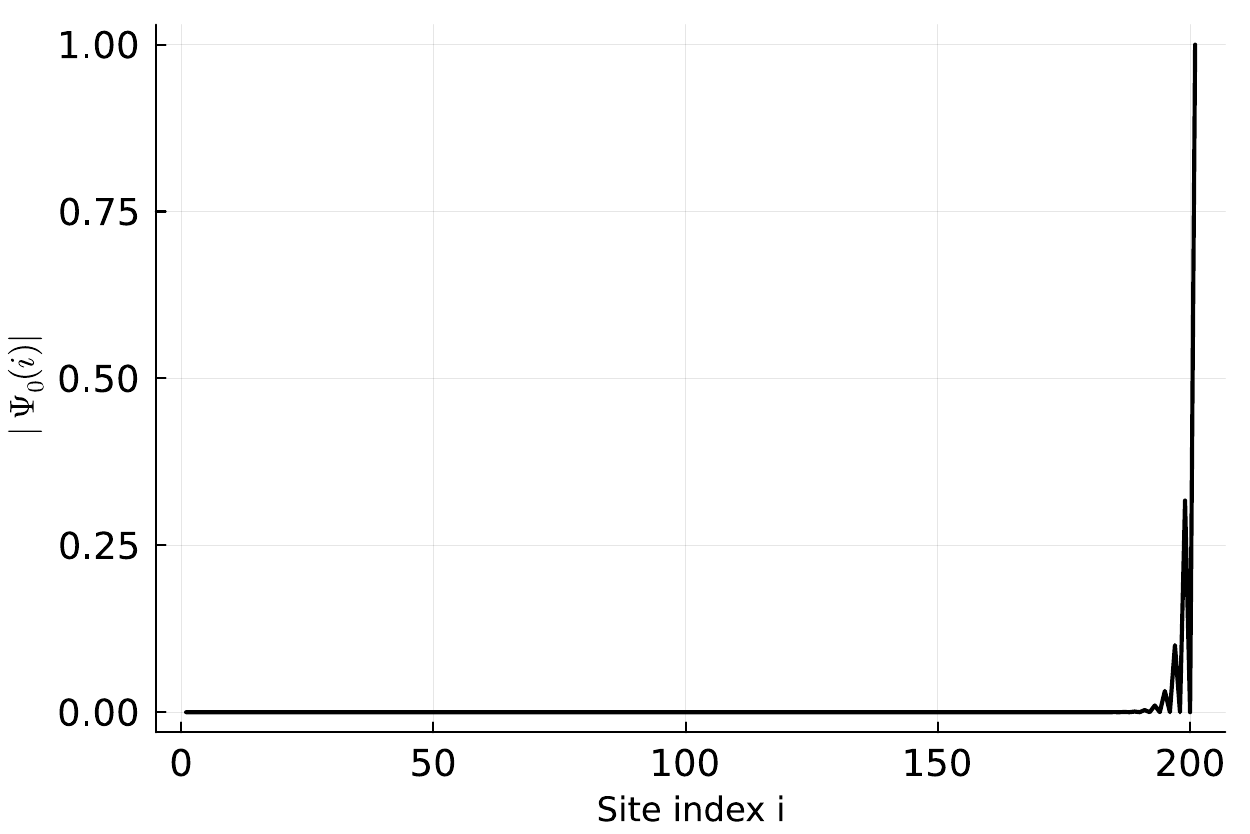}
     \caption{}
  \end{subfigure}
      \hspace{2pt}
  \begin{subfigure}[h]{0.32\textwidth}
     \includegraphics[width=\textwidth]{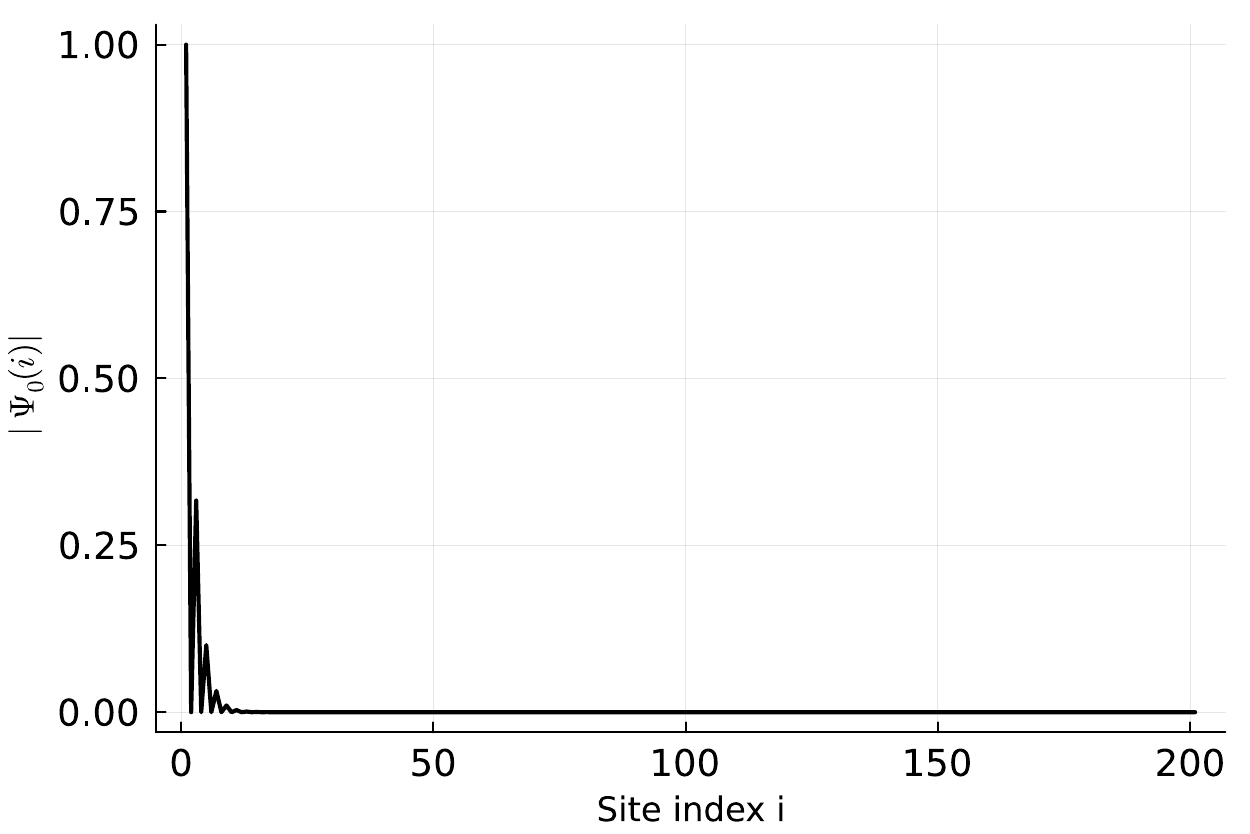}
     \caption{}
  \end{subfigure}
\caption{(a) The eigenspectrum of the complex SSH model (odd number of sites) with $v = 1 + i$ and $w = 1 + r \, i$, for $N = 201$. Since there always exists a zero energy state, there is always an edge state that accumulates at the end of the chain: (b) edge state at $w = 1 + 1.5 i$ and $v = 2 + 2.5 i$. (c) edge state at  $v = 1 + 1.5 i$ and $w = 2 + 2.5 i$\label{odd}}
\end{figure*}

\subsection{Eigenspectrum and the bulk-boundary correspondence of the complex SSH model}

In this section, we observe the behaviour of the complex SSH model for both even and odd number of sites. The analytical solution for the eigenspectrum and the expressions for the eigenvectors are given in Appendix \ref{analytical}.

The behaviour of the complex SSH model with even number of sites is expected. As can be seen in Fig.~\ref{fig:numresult} and Fig. \ref{odd}, the overall spectrum behaviour for the SSH model is similar to that of the original SSH model \cite{sirker2014boundary}. When the number of sites is even, the band gap closes where the topological phase transition happens. As observed in Fig.~\ref{fig:numresult}(a), when $|v| = |w|$, the band gap closes. After the band closes, edges states can be observed (see Fig.~\ref{fig:numresult}(b)). 

The topological phase transition points for the even case, where edge states start to occur, can also be predicted through the Bloch Hamiltonians alone. This shows clear signs of the existence of bulk-boundary correspondence. As shown in Fig.~\ref{fig:numresult}(c)-(e), the band gap closes at $|v| = |w|$, which is where the zero energy edge states begin to appear in the open system.

In comparison, for the odd case, the translational symmetry of the system is broken, and the bulk can no longer predict the boundary behaviour. There always exists a zero energy state and the corresponding edge state can always be observed (see Fig.~\ref{odd}). Interestingly, the band gap also closes at the same spot where the topological phase transition happens for the even case, and the behaviour of the edge states changes. The electron density accumulates always at the broken dimer at the end of the chain. When $|v| > |w|$, the peak occurs at the right hand side, while the peaks occurs on the left hand side when $|v| < |w|$. 

\section{Conclusion\label{conclusion}}

We have used two methods -- Berry phase and topological data analysis -- to explore the topological properties of the complex SSH model. Our findings demonstrate that both methods can generate topological phase diagrams for the model, identifying two distinct regions based on the relative magnitudes of the complex parameters $|v|$ and $|w|$. Specifically, we find that when $|v| > |w|$, the system is topologically trivial, whereas for $|v| < |w|$, it exhibits topologically non-trivial behavior. These results provide insights into the topological properties of the complex SSH model and highlight the complementary potential of the two methods in the topological analysis of the SSH-type models.

We have specifically focused on the Hermitian version of the complex SSH model as a first step towards testing and exploring the limitations and potential generalisations of topological approaches for the more complicated SSH models, including their non-Hermitian extensions. We recognise some potential challenges going to non-Hermitian systems. In terms of the Berry phase approach, the quantum geometry of the non-Hermitian system is no longer Riemannian. With the topological data analysis approach, preliminary work indicates that the geometry of the projected Hilbert space becomes more complex when the system deviates from the standard SSH model. This observation comes from several attempts using different definitions of fidelity \cite{ye2023quantum} in non-Hermitian systems. For future approaches using this method, it is suspected that more nuanced tools from topological data analysis will be needed to be able to distinguish among different topological phases.

%% file: appendix.tex
\section{Implementation for the detection of the winding number \label{windingprogram}}

\begin{figure*}
\centering
\includegraphics[width=0.25\textwidth,angle=270]{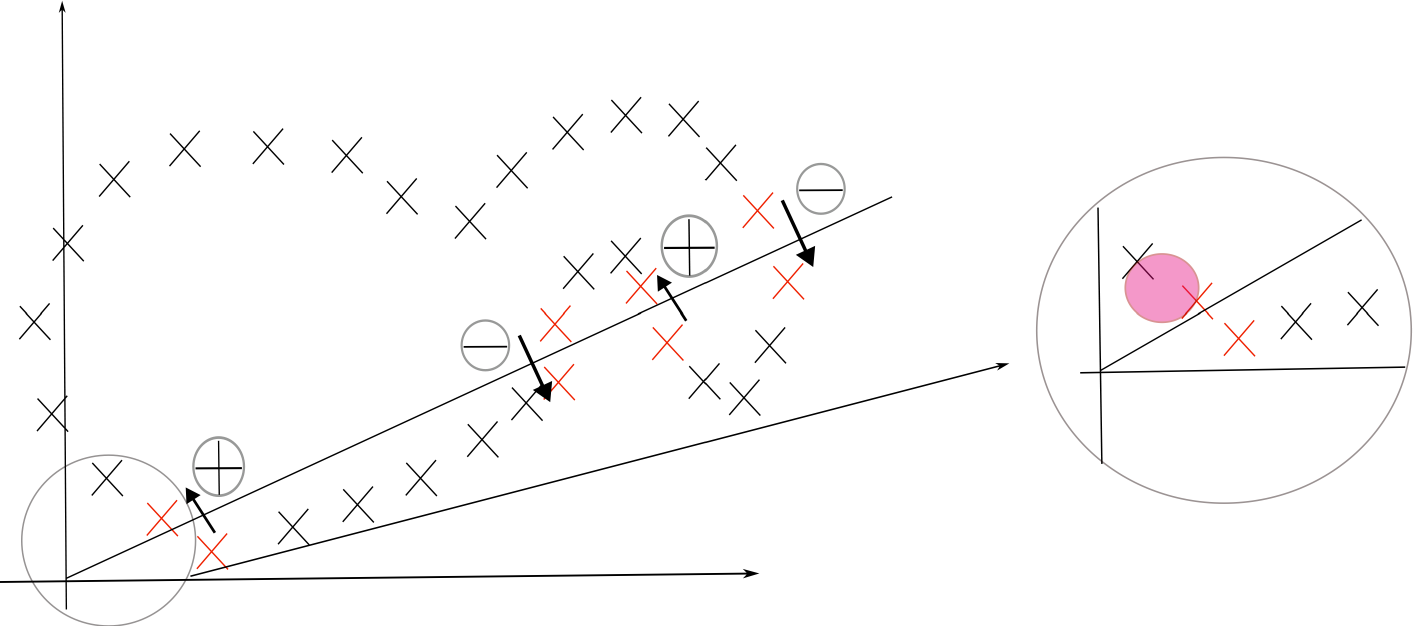}
\caption{The computational implementation of the winding number detection program.}
\label{winding_program}
\end{figure*}

The winding number, as the name suggests, is defined as the number of times a closed path winds around a given pole. It is a topological invariant as it captures the 1-homotopy class of $S^1$ ($\pi_1 (S^1) = Z$). Our program outputs the winding number with an ordered set of coordinates in $R^2$ and the poles as inputs. We calculate the winding number for one pole at a time by translating the pole to the origin.

The general idea of the program is to add up the orientation of the curve at points of intersection between the curve and a straight line from the origin, the pole (see Fig.~\ref{winding_program}). The idea is founded in degree theory \cite{brouwer1911abbildung}, and the implementation shares similarity with Ref.~\cite{franek2015effective}.

In terms of implementation, the orientation of the curve is provided by the order in the coordinates. The intersections are found by identifying the neighbouring points that ``sandwich" the straight line from the origin. Once the intersections are found, it is straightforward to sum up all the orientations at the intersections and outputs the winding number.

There are several minor challenges in this simple procedure. The first one comes from the collection of the points from the closed loop. The collection of points is important as too sparse sets would affect the outcome. The assumption is the sets of points gathered are sufficient to represent the local curvatures of the closed loop.

The second challenge is the detection of the poles sitting on the closed loop. The corresponding winding number should be undefined in this case. However, it is difficult to show the pole being on the curve when the curve is made up of discrete points. We provide the feature (can be toggled) of identifying the exceptional points by recognising if the pole is ``close enough" to two neighboring points. The pole is identified as close enough if it is inside the circle defined by the two closest neighbouring points as antipodal points.

In the case of the complex SSH model, the detection of the undefined winding number is less significant because of the nature of the phase diagram (see Fig.~\ref{Berry}). The topologically undefined region exists on the boundary of two phases and is therefore visually obvious. In other more complicated cases where a region can be dedicated to be topologically undefined, this feature would be more significant.
\section{Numerical details for step 2 \label{num detail}}

In the step PD1, Flux is the Julia package used to implement the optimisation of the sampling of $k$-space. Adam, which is an extension to the stochastic gradient descent algorithm, was chosen in Flux to optimise the sampling~\cite{innes:2018}. 

The package we used to calculate the persistence diagrams is a pure Julia implementation of the Ripser algorithm, Ripserer~\cite{cufar2020ripserer}.\

\section{Analytical solution for the complex SSH model \label{analytical}}

The tridiagonal matrix of size $N \times N$ in the basis $\ket{m,A/B}$ for the complex SSH Hamiltonian (\ref{SSH1}) is of the form 
\begin{center}
$H$ = 
$\begin{bmatrix} 
0& v & 0 & 0 & 0 & 0 & \dots\\
v^* & 0& w & 0 & 0 & 0 &\dots \\
0 & w^* & 0& v & 0 & 0 & \dots \\
0 & 0 & v^* & 0 & w & 0 & \dots \\
0 & 0 & 0 & w^* & 0 & v  & \dots\\
0 & 0 & 0 & 0 & v^* & 0 & \dots \\
\vdots & \hdots & \\
0 &      &      &   &     &   &  \hdots & 0
\end{bmatrix} $ \,.
\end{center}

\subsection{The case when $N$ is odd, $\lambda \not= 0$}

As noted in the solution of the SSH model with real $v$ and $w$~\cite{sirker2014boundary}, the diagonalisation of 
such a matrix has been discussed in the mathematics literature~\cite{shin1997formula}, from which we can make use of Lemma 1~\cite{shin1997formula} in terms of the definition for T, the Fibonacci-type sequence
\begin{equation}
T(\alpha,\beta,n) =  \alpha^n + \alpha^{n-1} \beta + \cdots + \alpha \beta^{n-1} + \beta^n \,.   
\end{equation}

\textbf{Lemma 1}. Let $\alpha$ and $\beta$ be nonzero complex numbers. Then the Fibonacci-type sequence $\{a_n\}^{\infty}_{n=1}$ defined as
\begin{align}
    a_{n+2} - (\alpha + \beta) a_{n+1} + \alpha \beta a_n = 0,~~ n = 1,2,\ldots
\end{align}
can be represented in terms of $a_1$ and $a_2$ by
\begin{align}
a_n = T(\alpha,\beta,n-2)a_2 - \alpha \beta T(\alpha,\beta,n-3)a_1, ~~n=1,2,\ldots
 \end{align}

Following \cite{shin1997formula}, if $N = 2M + 1$, the eigenvector components satisfy the set of equations 
\begin{align} 
    v x_2 &= \lambda x_1, \label{a} \\
    w^* x_{2M} &= \lambda x_{2M+1} , \label{b}\\
    v^{*} x_{2n-1} + w x_{2n+1} &= \lambda x_{2n} , \label{c}\\
    w^{*} x_{2n} + v x_{2n+2} &= \lambda x_{2n+1} , \label{d}
\end{align}
where $\lambda$ are the eigenenergies.

From equations (\ref{c}) and (\ref{d}), we have
\begin{align}
    x_{2n+3} + \frac{vv^* + ww^* - \lambda^2}{vw} x_{2n+1} + \frac{w^*v^*}{wv} x_{2n-1} &= 0. \\
     x_{2n+2} + \frac{vv^* + ww^* - \lambda^2}{vw} x_{2n} + \frac{w^*v^*}{wv} x_{2n-2} &= 0.
\end{align}
Therefore, we can choose $\alpha$ and $\beta$ such that
\begin{align}
    \alpha + \beta &= \frac{vv^* + ww^* - \lambda^2}{vw} x_{2n+1},\\
    \alpha \beta & = \frac{w^*v^*}{wv}.
\end{align}
From equations (\ref{a})-(\ref{d}), we have
\begin{align} \label{4}
    x_2 &= \frac{\lambda}{v} x_1, \\
    x_3 &= \left[\frac{v(\alpha + \beta) + w^*}{v}\right]x_1, \\
    x_4 &= \frac{\lambda (\alpha + \beta)}{v} x_1 .
\end{align}
Combining Lemma 1 with these equations then gives
\begin{align}
    x_{2n-1} &= \left[\frac{w^*}{v} T(\alpha,\beta,n-2) + T(\alpha,\beta,n-1)\right]x_1 \label{oddeigenvec1}, \\
    x_{2n} &= \frac{\lambda}{v} T(\alpha,\beta,n-1)x_1 .\label{oddeigenvec2}
\end{align}
Now combining this result with equation (\ref{b}) gives
\begin{align}
    T(\alpha,\beta,N) = 0 \,.
\end{align}
Therefore we have
\begin{align} \label{0}
    \alpha^{M+1} - \beta^{M+1} &= 0 \,,\\
    |\alpha| &= |\beta|\,.
\end{align}
A natural choice for $\alpha$ and $\beta$ follows as 
\begin{equation}
    \alpha = \frac{w^*}{w} \exp{ki} \,, \quad 
    \beta = \frac{v^*}{v} \exp{-ki} \,. \label{alphabeta}
    \end{equation}
Substitution into (\ref{0}) then gives
\begin{align}\label{k}
    \left(\frac{w^*}{w}\right)^{M+1} &\exp{ik(M+1)}\nonumber\\ - \left(\frac{v^*}{v}\right)^{M+1}& \exp{-ik(M+1)} = 0 \, .
\end{align}
When $\frac{w^*}{w} = \frac{v^*}{v}$, the solution reduces to 
\begin{align}
    k = \frac{n\pi}{M+1}, \quad n \in [1,M] \,. 
\end{align}

In summary, we have the eigenvalues in the form
\begin{align} \label{eigenvalue}
    \lambda = \pm \sqrt{|v|^2 + |w|^2 + vw(\alpha + \beta)} \,,
\end{align}
where $\alpha$ and $\beta$ are defined in (\ref{alphabeta}) and $k$ needs to be solved from equation (\ref{k}). 
The eigenvector components are given by equations (\ref{oddeigenvec1}) and (\ref{oddeigenvec2}).

\subsection{The case when $N$ is even, $\lambda \not= 0$}

Again following Ref.~\cite{shin1997formula}, if $N = 2M$, we have the set of equations
\begin{align} 
    v x_2 &= \lambda x_1 \,,\label{ae} \\
    v^* x_{2M-1} &= \lambda x_{2M} \,, \label{be}\\
    v^{*} x_{2n-1} + w x_{2n+1} &= \lambda x_{2n} \,,\label{ce}\\
    w^{*} x_{2n} + v x_{2n+2} &= \lambda x_{2n+1} \,. \label{de}
\end{align}
Combining these equations with   (\ref{oddeigenvec1}) and (\ref{oddeigenvec2}) gives
\begin{align}\label{even}
    \frac{v^* w^*}{v} T(\alpha,\beta, M-2) + v^* T(\alpha,\beta,M-1) & \nonumber\\
    = \frac{\lambda^2}{v} T(\alpha,\beta,M-1) \,.
\end{align}

The eigenvalue expression is the same as equation (\ref{eigenvalue}). The eigenvectors are as given in (\ref{oddeigenvec1}) and (\ref{oddeigenvec2}), where $k$ needs to be solved using the anstaz in (\ref{alphabeta}) in combination with (\ref{even}), which can be readily done.

\subsection{$\lambda = 0$}

From equations (\ref{a}), (\ref{c}) and (\ref{d}), the form of the eigenvectors for the zero energy edge states is
\begin{align}
    x_{2n+1} &= \left(-\frac{v^*}{w}\right)^n x_{2n-1} \,,\\
    x_{2n} & = 0 \,.
\end{align}